\documentclass[aps,prb,a4paper,twocolumn,floatfix,showkeys,amsmath,amssymb,nofootinbib,nobibnotes,altaffilletter]{revtex4}

\usepackage[T1]{fontenc}
\usepackage[latin1,utf8]{inputenc}
\usepackage[english]{babel}

\usepackage{setspace}

\usepackage{amsmath}
\usepackage{bm}

\usepackage{graphicx}
\usepackage{pslatex}
\usepackage{xspace}
\usepackage{subfigure}
\usepackage{float}

\usepackage{booktabs}

\usepackage{ragged2e}
\usepackage[labelfont=bf, format=plain, indention=0.5cm, textfont=small, skip=4pt, justification=RaggedRight, tableposition=top, figureposition=bottom ]{caption}

\newcommand{\degree}{$^{\circ}$}

\renewcommand{\eqref}[1]{Eq.~(\ref{eq:#1})}

\begin{document}

\preprint{}

\title{Photoinduced C$_{\textup{70}}$ radical anions in polymer:fullerene blends}

\author{Andreas \surname{Sperlich}}\email{sperlich@physik.uni-wuerzburg.de}
\author{Julia \surname{Kern}}
\author{Hannes \surname{Kraus}}
\author{Carsten \surname{Deibel}}
\affiliation{Experimental Physics VI, Julius-Maximilians-University of W{\"u}rzburg, 97074 W{\"u}rzburg, Germany}

\author{Vladimir \surname{Dyakonov}}
\author{Moritz \surname{Liedtke}}
\affiliation{Experimental Physics VI, Julius-Maximilians-University of W{\"u}rzburg, 97074 W{\"u}rzburg, Germany}
\affiliation{Bavarian Centre for Applied Energy Research (ZAE Bayern), 97074 W{\"u}rzburg, Germany}

\author{Salvatore \surname{Filippone}}
\affiliation{Departamento de Qu\'imica Org\'anica, Universidad Complutense de Madrid, 28040 Madrid, Spain}

\author{Juan Luis \surname{Delgado}}
\author{Nazario \surname{Mart\'in}}
\affiliation{Departamento de Qu\'imica Org\'anica, Universidad Complutense de Madrid, 28040 Madrid, Spain}
\affiliation{IMDEA-Nanociencia, Facultad de Ciencias, Ciudad Universitaria de Cantoblanco, 28049 Madrid, Spain}

\date{February 8, 2011}


\begin{abstract}
Photoinduced polarons in solid films of polymer--fullerene blends were studied by photoluminescence (PL), photoinduced absorption (PIA) and electron spin resonance (ESR). The donor materials used were P3HT and MEH-PPV. As acceptors we employed PC$_{\textup{60}}$BM as reference and various soluble C$_{\textup{70}}$-derivates: PC$_{\textup{70}}$BM, two different diphenylmethano-[70]fullerene oligoether (C$_{\textup{70}}$-DPM-OE) and two dimers, C$_{\textup{70}}$--C$_{\textup{70}}$ and C$_{\textup{60}}$--C$_{\textup{70}}$. Blend films containing C$_{\textup{70}}$ revealed characteristic spectroscopic signatures not seen with C$_{\textup{60}}$. Light-induced ESR showed signals at g$\geq$2.005, assigned to an electron localized on the C$_{\textup{70}}$ cage. The formation of C$_{\textup{70}}$ radical anions also leads to a  subgap PIA band at 0.92~eV, hidden in the spectra of C$_{\textup{70}}$-based P3HT and MEH-PPV blends, which allows for more exact studies of charge separated states in conjugated polymer:C$_{\textup{70}}$ blends.
\end{abstract}



\maketitle

\begin{figure}[h]
  \centering
  \includegraphics[width=\columnwidth]{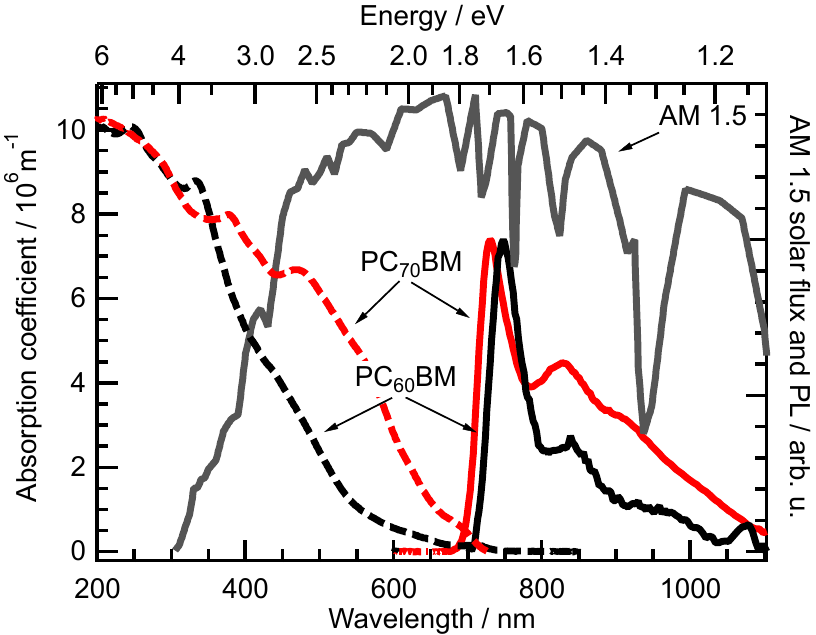}
  \caption{AM 1.5 solar flux (normalized), absorption at 300~K (dashed) and normalized PL at 12~K (solid) of thin films of PC$_{\textup{60}}$BM (black) and PC$_{\textup{70}}$BM (red).}
  \label{fgr:absPL}
\end{figure}

Due to their potentially low manufacturing costs, thin film organic solar cells (OSC) may become highly competitive in the area of direct solar energy conversion. Efficiencies of  8.3~\% have already been reported \cite{green2011,deibel2010review}.
For directed optimization efforts a better understanding of the photophysical processes in the device during the conversion of incident light into electrical energy is required. Using blends of conjugated polymers and fullerenes, an efficient photoinduced charge separation can be achieved, subsequently followed by migration of the charges to the device electrodes.
PC$_{\textup{60}}$BM, a soluble fullerene derivative, is used as acceptor material in the vast majority of the OSC devices reported. It was outperformed recently by PC$_{\textup{70}}$BM, due to its higher absorption coefficient in the visible part of the solar spectrum (figure~\ref{fgr:absPL}), but also due to the slightly higher open circuit voltage \cite{wienk2003,liang2010}.

Although understanding of the elementary steps of efficient charge separation in the photovoltaic materials is a prerequisite for improving the efficiency, the photophysics of C$_{\textup{70}}$~containing blends is much less investigated than its C$_{\textup{60}}$ relatives. 
So far ESR and VIS-NIR spectra of C$_{\textup{70}}$~derivates obtained in different phases like crystallites, salt-like samples and dilute (frozen) solutions in (un)polar solvents were previously presented in the literature \cite{baumgarten1996,hase1995,fujitsuka1997,konarev2003}. The resulting signals differ for each different phase, due to interaction with the host material. Therefore from the ESR measurements the undisturbed g-tensor components of the C$_{\textup{70}}$ radical anion could not be obtained. Additionally several partly overlapping NIR-absorption bands of single and multiply charged C$_{\textup{70}}$ anions and cations were detected. An absorption at 1370~nm (0.905~eV) was tentatively assigned to solvation complexes of C$_{\textup{70}}$ radical anions in solution \cite{baumgarten1996,fujitsuka1997}. 
Recently, we reported on new C$_{\textup{70}}$-based fullerene dimers showing additional PIA and ESR features when blended with conjugated polymers in thin solid films \cite{delgado2009}. We investigated the symmetry of the resulting g-tensor in polymer:C$_{\textup{70}}$ blends \cite{poluektov2010}. Here we generalize these findings for a broader class of C$_{\textup{70}}$ fullerene derivates in solid state and present results of the subgap PIA determining the excitation energies of the involved excited state.

\begin{figure}[h]
\centering
  \includegraphics[width=\columnwidth]{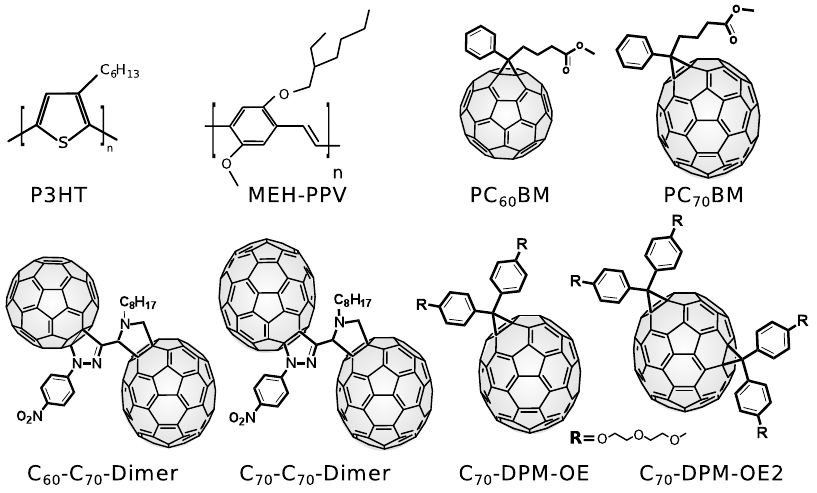}
  \caption{Donor-polymers P3HT and MEH-PPV. Employed acceptors were C$_{\textup{70}}$ fullerene derivates and PC$_{\textup{60}}$BM  as reference.}
  \label{fgr:struc}
\end{figure}

P3HT (Rieke Metals), MEH-PPV (Sigma-Aldrich), PC$_{\textup{60}}$BM and PC$_{\textup{70}}$BM (Solenne B.V.) were used as purchased. Details about the synthesis of the fullerene dimers and C$_{\textup{70}}$-DPM-OE are reported elsewhere \cite{delgado2009}.
Samples were processed from chlorobenzene solutions (10~mg/ml) with a 1:1 weight ratio for blends. Thin films of $\sim$100~nm were spincast onto saphire substrates for use in PIA, PL and absorption. For ESR 200~$\mu l$ solution were vacuum-dried inside an ESR sample tube. All samples were annealed afterwards for 10~min at 130~\degree C.

Figure~\ref{fgr:absPL} shows that PC$_{\textup{70}}$BM has an increased absorption in the range of 400--700~nm compared to PC$_{\textup{60}}$BM, which is favorable in view of the AM~1.5 solar flux. Absorption was measured with a Perkin-Elmer Lambda 900 spectrometer under ambient conditions. At low temperatures (12~K) these samples emitted a weak, yet detectable PL when excited by a 532~nm laser. Intensities of the PL spectra were corrected for the spectral response of the spectrometer. The spectra of both fullerene derivatives showed well resolved vibronic structures and the PL of PC$_{\textup{70}}$BM was blue-shifted by $\sim$40~meV (17~nm) matching literature values~\cite{sibley1992}.

\begin{figure}[h]
  \centering
  \includegraphics[width=\columnwidth]{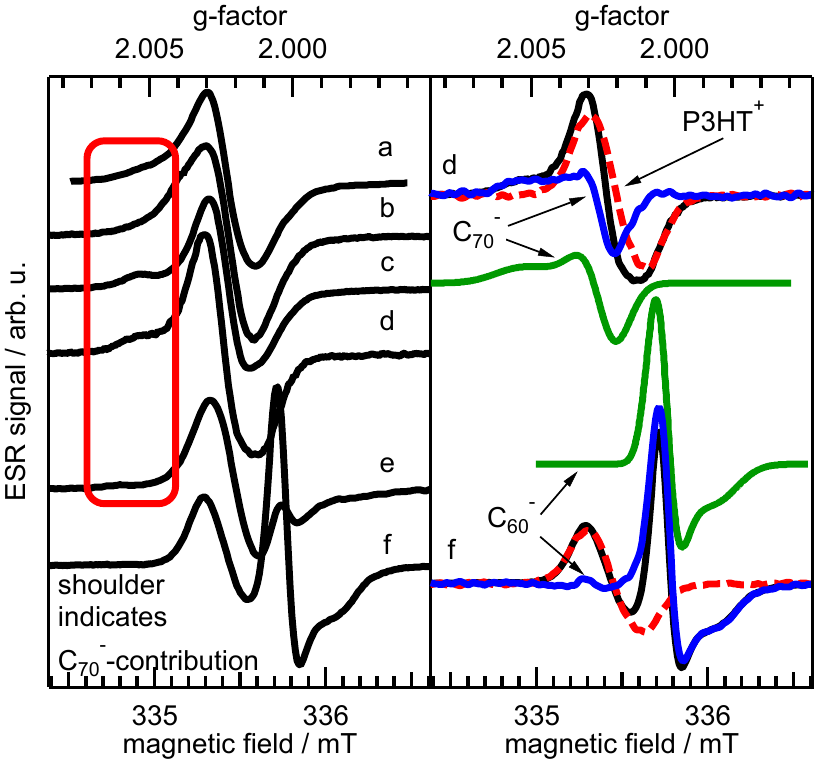}
  \caption{~\textbf{Left side:} ESR spectra at 100~K of P3HT blended with different fullerenes: C$_{\textup{70}}$--C$_{\textup{70}}$-Dimer (a), C$_{\textup{70}}$-DPM-OE2 (b), C$_{\textup{70}}$-DPM-OE (c), PC$_{\textup{70}}$BM (d), C$_{\textup{60}}$--C$_{\textup{70}}$-Dimer (e), PC$_{\textup{60}}$BM (f). All C$_{\textup{70}}$-composites show a shoulder for g$\geq$2.005 (indicated by the red rectangle) which is assigned to the C$_{\textup{70}}$ anion. \textbf{Right side:} Deconvolution of the experimental spectra (black), the positive polaron P3HT$^{+}$ (dashed red), the fullerene anions (blue) and their simulated spectra (green) for composites (d) and (f).}
  \label{fgr:ESR}
\end{figure}

In figure~\ref{fgr:ESR} (left) the light-induced ESR spectra (normalized) of P3HT blended with the different fullerene derivatives are shown. Illumination was provided by a 532~nm laser and the samples were cooled to 100~K. All spectra inhibit the typical derivative-like absorption line at g=2.002, originating from positive polarons on the polymer chain. In P3HT:PC$_{\textup{60}}$BM blends, the radical anion of C$_{\textup{60}}$ can be detected at g=2.000. For blends consisting of P3HT and C$_{\textup{70}}$-derivatives no such signal can be identified directly, as the C$_{\textup{70}}$ radical anion line is strongly superimposed with the polymer polaron signal. Instead all C$_{\textup{70}}$-containing blends disclose a shoulder at g$\geq$2.005, being absent in the spectrum of P3HT:PC$_{\textup{60}}$BM. In figure~\ref{fgr:ESR} (right), the deconvolution of the experimental spectra into their contributions of the positive polymer polaron, the C$_{\textup{70}}$ (top) and  C$_{\textup{60}}$ (bottom) radical anion are shown. The deconvolution method has been described elsewhere \cite{poluektov2010}. We were able to simulate the spectra of the fullerene-anions (using Simfonia) with the following g-tensor components $(g_{xx} / g_{yy} / g_{zz})$: C$_{60}^{-}= (2.00058 / 2.00045 / 1.99845)$ and C$_{70}^{-}= (2.00592 / 2.00277 / 2.00211)$. A closer look at the spectrum of P3HT:C$_{\textup{60}}$--C$_{\textup{70}}$-dimer (e) revealed, that contributions of both fullerene anions are present, hinting that the charge transfer process is not dominated by one fullerene.

\begin{figure}[h]
  \centering
  \includegraphics[width=\columnwidth]{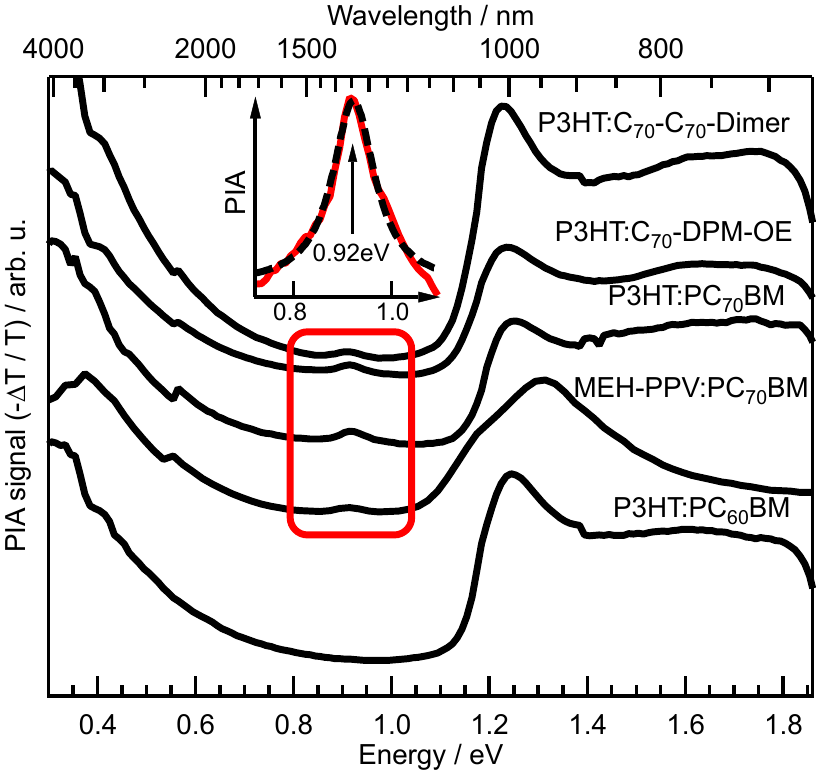}
  \caption{PIA spectra at 15--30~K. While the structure of the spectra is similar, the peak at 0.92~eV shows up only in blends with C$_{\textup{70}}$-derivates.  \textbf{Inset}: Subtraction (red) of the normalized spectra of P3HT:PC$_{\textup{60}}$BM from P3HT:PC$_{\textup{70}}$BM together with a lorentzian fit of the lineshape (dashed black).}
  \label{fgr:PIA}
\end{figure}

Further, we studied the photophysical properties by photoinduced absorption (PIA). The excitation by a 532~nm laser leads to singlet excitons, which are separated at the polymer:fullerene interface yielding positive and negative charge carriers in the polymer and the fullerene phase, respectively. The resulting change in transmission of a white light continuum probe beam was measured ($-\Delta T / T$). The transmitted light was focussed into a monochromator and the detection was provided by a silicon photodiode (550--1030~nm) and a liquid nitrogen cooled InSb-detector (1030--5500~nm). In most polymer:fullerene blends, transitions ascribed to polaronic states (radical cations) are located at 0.3~eV and 1.2~eV. For P3HT or PPV blended with PC$_{\textup{60}}$BM these spectral signatures are well documented \cite{osterbacka2000,montanari2002}. 

In the PIA spectra of all polymer:C$_{\textup{70}}$ blends, similar polaronic absorption bands could be detected, indicating efficient charge separation. Additionally we found an absorption at 0.92~eV (figure~\ref{fgr:PIA}), which is absent in the reference P3HT:PC$_{\textup{60}}$BM spectrum and therefore related to the C$_{\textup{70}}$. As the examined fullerene derivatives have different side chains, we can exclude electrons being localized on here. 
Charge transfer states as possible candidates for a new subgap absorption could be ruled out by replacing the host polymer P3HT with MEH-PPV, which has different HOMO-LUMO energies \cite{holt2005,oku2008}.
Furthermore the relation of the 0.92~eV peak to C$_{\textup{70}}$-derivatives is also consistent with the different temperature dependence of the PIA peaks in the blend of MEH-PPV:PC$_{\textup{70}}$BM (data not shown). The polaron peaks at 0.3~eV and 1.2~eV show both an identical decrease in signal intensity when increasing the temperature from 30~K to 300~K, while the 0.92~eV peak decreases steeper and vanishes at 300~K. 
We were able to separate the contribution of the underlying peak by subtracting the normalized spectrum of P3HT:PC$_{\textup{60}}$BM from P3HT:PC$_{\textup{70}}$BM . This is shown together with a Lorentzian fit of the spectrum in the inset in figure~\ref{fgr:PIA} (E$_{0}=$~0.92~eV / 1350~nm and FWHM~$=$~0.11~eV).

To summarize, we have studied a variety of solid films of C$_{\textup{70}}$-fullerene derivates and dimers blended with two different conjugated polymers, with the emphasis on the photogenerated radical anion. We identified spectral features originating from C$_{\textup{70}}$ in PIA and ESR spectroscopy, namely an additional subgap PIA peak at 0.92~eV and an ESR-shoulder at g$\geq$2.005. By comparing results from different blends we can infer that these features are indeed signatures of the C$_{\textup{70}}$ radical anion and not originating from the side chains of the fullerenes or charge transfer states at polymer:C$_{\textup{70}}$-fullerene interfaces. As anticipated, ESR signatures related to the C$_{\textup{70}}$ radical anion were found and the corresponding spectra could be simulated with the g-values from our previous findings \cite{poluektov2010}. The obtained signatures are of importance for studying the charge transfer reactions in C$_{\textup{70}}$-based bulk-heterojunctions.


\begin{acknowledgments}
The work at the University of W\"urzburg was supported by the German Research Foundation, DFG, within the SPP 1355 ``Elementary processes in organic photovoltaics'' (DY18/6-1 and -2). The MICINN of Spain (project CT2008-00795/BQU, R\&C program, and Consolider-Ingenio 2010C-07-25200) and the CAM (project P-PPQ-000225-0505) are also acknowledged. J. L. D. thanks MICINN of Spain for a Ramon y Cajal contract cofinanced by the EU social funds.
\end{acknowledgments}

\providecommand{\WileyBibTextsc}{}
\let\textsc\WileyBibTextsc
\providecommand{\othercit}{}
\providecommand{\jr}[1]{#1}
\providecommand{\etal}{~et~al.}


\begin{thebibliography}{[10]}

\bibitem{green2011}
 \textsc{M.\,A. Green},  \textsc{K.~Emery},  \textsc{Y.~Hishikawa},  and
  \textsc{W.~Warta},
 \jr{Prog. Photovolt: Res. Appl} \textbf{19}(1), 84 (2011).


\bibitem{deibel2010review}
 \textsc{C.~Deibel} and  \textsc{V.~Dyakonov},
 \jr{Rep. Prog. Phys.} \textbf{73}(16), 096401 (2010).


\bibitem{wienk2003}
 \textsc{M.~Wienk},  \textsc{J.\,M. Kroon},  \textsc{W.\,J.\,H. Verhees},
  \textsc{J.~Knol},  \textsc{J.\,C. Hummelen},  \textsc{P.\,A. van Hal},  and
  \textsc{R.~Janssen},
 \jr{Angew. Chem. Int. Ed.} \textbf{42}(29), 3371 (2003).


\bibitem{liang2010}
 \textsc{Y.~Liang},  \textsc{Z.~Xu},  \textsc{J.~Xia},  \textsc{S.\,T. Tsai},
  \textsc{Y.~Wu},  \textsc{G.~Li},  \textsc{C.~Ray},  and
  \textsc{L.~Yu},
 \jr{Adv. Mater.} \textbf{22}(20), E135 (2010).


\bibitem{baumgarten1996}
 \textsc{M.~Baumgarten} and  \textsc{L.~Gherghel},
 \jr{Appl. Magn. Reson.} \textbf{11}(2), 171 (1996).


\bibitem{hase1995}
 \textsc{H.~Hase} and  \textsc{Y.~Miyatake},
 \jr{Chem. Phys. Lett.} \textbf{245}, 95 (1995).


\bibitem{fujitsuka1997}
 \textsc{M.~Fujitsuka},  \textsc{A.~Watanabe},  \textsc{O.~Ito},
  \textsc{K.~Yamamoto},  and  \textsc{H.~Funasaka},
 \jr{J. Phys. Chem. A} \textbf{101}(43), 7960 (1997).


\bibitem{konarev2003}
 \textsc{D.\,V. Konarev},  \textsc{S.\,S. Khasanov},  \textsc{G.~Saito},
  \textsc{A.~Otsuka},  \textsc{Y.~Yoshida},  and  \textsc{R.\,N.
  Lyubovskaya},
 \jr{J. Am. Chem. Soc.} \textbf{125}(33), 10074 (2003).


\bibitem{delgado2009}
 \textsc{J.\,L. Delgado},  \textsc{E.~Esp\'ildora},  \textsc{M.~Liedtke},
  \textsc{A.~Sperlich},  \textsc{D.~Rauh},  \textsc{A.~Baumann},
  \textsc{C.~Deibel},  \textsc{V.~Dyakonov},  and  \textsc{N.~Mart\'in},
 \jr{Chem. Eur. J.} \textbf{15}(48), 13474 (2009).


\bibitem{poluektov2010}
 \textsc{O.\,G. Poluektov},  \textsc{S.~Filippone},  \textsc{N.~Mart\'in},
  \textsc{A.~Sperlich},  \textsc{C.~Deibel},  and  \textsc{V.~Dyakonov},
 \jr{J. Phys. Chem. B} \textbf{114}(45), 14426 (2010).


\bibitem{sibley1992}
 \textsc{S.\,P. Sibley},  \textsc{S.\,M. Argentine},  and  \textsc{A.\,H.
  Francis},
 \jr{Chem. Phys. Lett.} \textbf{188}(3), 187 (1992).


\bibitem{osterbacka2000}
 \textsc{R.~\"Osterbacka},  \textsc{C.\,P. An},  \textsc{X.\,M. Jiang},  and
  \textsc{Z.\,V. Vardeny},
 \jr{Science} \textbf{287}(5454), 839 (2000).


\bibitem{montanari2002}
 \textsc{I.~Montanari},  \textsc{A.\,F. Nogueira},  \textsc{J.~Nelson},
  \textsc{J.\,R. Durrant},  \textsc{C.~Winder},  \textsc{M.\,A. Loi},
  \textsc{N.\,S. Sariciftci},  and  \textsc{C.~Brabec},
 \jr{Appl. Phys. Lett.} \textbf{81}(16), 3001 (2002).


\bibitem{holt2005}
 \textsc{A.\,L. Holt},  \textsc{J.\,M. Leger},  and  \textsc{S.\,A.
  Carter},
 \jr{J. Chem. Phys.} \textbf{123}(4), 044704 (2005).


\bibitem{oku2008}
 \textsc{T.~Oku},  \textsc{S.~Nagaoka},  \textsc{A.~Suzuki},
  \textsc{K.~Kikuchi},  \textsc{Y.~Hayashi},  \textsc{H.~Inukai},
  \textsc{H.~Sakuragi},  and  \textsc{T.~Soga},
 \jr{Journal of Physics and Chemistry of Solids} \textbf{69}(5), 1276 (2008).


\end{thebibliography}
\end{document}